%% file: main.tex
\documentclass[sigconf]{acmart}
\usepackage{bm}
\usepackage{array}
\usepackage{graphicx} 
\usepackage{booktabs,multirow,adjustbox}
\graphicspath{{Figures/}}      
\DeclareGraphicsExtensions{.pdf,.png,.jpg,.tex}
\usepackage{xcolor}
\definecolor{dip}{HTML}{D62F2F}

\usepackage{forest}
\begin{document}


\title{A Taxonomy-Driven Case Study of Australian Web Resources Against Technology-Facilitated Abuse}


\author{Dipankar Srirag}
\email{d.srirag@unsw.edu.au}
\affiliation{%
  \institution{University of New South Wales}
  \city{Sydney}
  \country{Australia}
}

 \author{Xiaolin Cen}
\email{xiaolin.cen@student.unsw.edu.au}
\affiliation{%
  \institution{University of New South Wales}
  \city{Sydney}
  \country{Australia}
}

 \author{Rahat Masood}
 \email{rahat.masood@unsw.edu.au}
\affiliation{%
  \institution{University of New South Wales}
  \city{Sydney}
  \country{Australia}
}

\author{Aditya Joshi}
\email{aditya.joshi@unsw.edu.au}
\affiliation{%
  \institution{University of New South Wales}
  \city{Sydney}
  \country{Australia}
}

\renewcommand{\shortauthors}{Srirag et al.}

\begin{abstract}
Technology-Facilitated Abuse (TFA) encompasses a broad and rapidly evolving set of behaviours in which digital systems are used to harass, monitor, threaten, or control individuals. Although prior research has documented many forms of TFA, there is no consolidated framework for understanding how abuse types, prevention measures, detection mechanisms, and support pathways relate across the abuse life cycle. This paper contributes (i) a unified, literature-derived taxonomy of TFA grounded in a structured review of peer-reviewed studies, and (ii) the first large-scale, taxonomy-aligned audit of institutional web resources in Australia. We crawl 306 government, non-government, and service-provider domains, obtaining 52,605 pages, and classify each using hierarchical zero-shot models to map web content onto our taxonomy. Complementary emotion and readability analyses reveal how institutions frame TFA and how accessible their guidance is to the public. Our findings show that institutional websites cover only a narrow subset of harms emphasised in the literature, with approximately 70\% of all abuse labelled pages focused on harassment, comments abuse, or sexual abuse, while less than 1\% address covert surveillance, economic abuse, or long-term controlling behaviours. Support pathways are similarly limited, with most resources (62\%) centred on digital information hubs rather than counselling, legal assistance, or community-based services. Readability analysis further shows that much of this content is written at late secondary or early tertiary reading levels (for example, ARI 18 to 22 across major categories), which may be inaccessible to a substantial portion of at-risk users. By highlighting strengths and gaps in Australia’s support for TFA, our taxonomy and audit method provide a scalable basis for evaluating institutional communication, improving survivor resources, and guiding safer digital ecosystems across the Australian context. The taxonomy itself serves as a solid foundation for analyses in other national contexts to foster awareness regarding technology-facilitated abuse.
\end{abstract}

\ccsdesc[500]{Security and privacy~Human and societal aspects of security and privacy}
\keywords{Technology-Facilitated Abuse, Cyber Abuse, Digital Safety, Digital Harms, Human-Centred Security}

\maketitle
\input{sections/01-intro}

\input{sections/02-tfa-taxonomy}
\input{sections/03-case-study}
\input{sections/04-conclusion}

\bibliographystyle{ACM-Reference-Format}
\bibliography{sample-base} 

\appendix

\section{Top Co-occurring Subcategories Across the TFA Taxonomies}
\label{sec:cooccurence}

Table~\ref{tab:tfa-cooccurrence} reports the most frequent subcategories for each primary taxonomy dimension. For TFA types, high counts for \textit{Public Shaming}, \textit{Threats of Eviction}, and \textit{Recorded Sexual Assault} indicate that institutional content largely focuses on overt, visible harms. In the prevention and detection taxonomies, subcategories such as \textit{Criminal Sanctions}, \textit{Family-Based Early Detection}, and \textit{In-Person Professional Assessment} show that responses are commonly framed through legal and clinical interventions rather than technical safeguards or community-level mechanisms. For support, the prominence of \textit{Online Support Communities} relative to categories such as \textit{Security Clinics} or \textit{Survivors Assistance Program} describes the central role of general digital information hubs compared to more intensive, programmatic services.

\input{tables/co-occurence}

\section{Example Domains Providing TFA Support Services}
\label{sec:examples}

Table~\ref{tab:support-examples} lists example Australian domains associated with each support category, as identified by the subcategory classification. Sites mapped to \textit{Digital Support Infrastructure} are typically information portals and resource hubs, while those associated with \textit{Technology Security Support} tend to provide device and account safety guidance. Domains linked to \textit{Institutional Support}, \textit{Psychological Support}, and \textit{Informal Network Support} are fewer and usually describe programs or referral options at a high level, reflecting the limited visibility of counselling, legal aid, and community-based support within the broader corpus.

\input{tables/support-examples}





\end{document}

%% file: sections/01-intro.tex
\section{Introduction}
\label{sec:intro}

Technology-Facilitated Abuse (TFA) refers to the use of networked technologies such as smartphones, social media platforms, online services, and smart devices to harass, monitor, threaten, or control another person~\cite{flynn2022technology, brown2021technology}. While many abusive behaviours predate digital media, web-facilitated infrastructures have intensified their reach and impact by enabling constant connectivity, anonymous or pseudonymous contact, and persistent traces that are difficult for victims to manage or erase~\cite{mckay2021standing, almansoori2024web, turk2024stop}. Emerging technologies such as Internet of Things (IoT) devices, location tracking, and generative media further expand the ways in which abuse can be enacted and concealed~\cite{abhinaya2024enabling, heinrich2024please, stephenson2023abuse, stephenson2023s}. These developments pose direct challenges to digital well-being and mental health~\cite{obada2022sok, o2023short, scott2023trauma} and undermine the UN Sustainable Development goals of gender equality and reduced inequalities~\cite{brown2021technology, henry2020technology}.

Existing research has documented a wide range of TFA scenarios, including intimate partner surveillance~\cite{almansoori2024web, ceccio2023sneaky}, image-based abuse~\cite{wei2024understanding, brown2021technology, flynn2022deepfakes, qin2024did}, financial sextortion~\cite{o2023short, liggett2025suicidal}, and online harassment~\cite{flynn2022technology, sheikh2024technology, abhinaya2024enabling, qin2024did, stephenson2023s, heinrich2024please}. Most studies rely on surveys~\cite{flynn2022technology, turk2024stop, flynn2021technology, brown2021technology, almansoori2024web, flynn2024workplace, liggett2025suicidal}, interviews~\cite{freed2019my, chen2022trauma, bellini2023paying, bellini2024abusive, gupta2024really, coduto2024delete, abhinaya2024enabling, qin2024did, heinrich2022airguard, stephenson2023s, bellini2023digital, dwyer2025friend}, or small scale qualitative analyses~\cite{mckay2021standing, stephenson2023abuse, wei2024understanding, wei2022anti, o2023short, wang2025breaking, chowdhury2024safe, pace2023every, stephenson2024sharenting} focused on specific populations or platforms. These approaches provide rich insight into lived experience, but they often lack a systematic way to compare different forms of TFA abuse and the kinds of resources that are supposed to prevent, detect, or respond to them. In particular, there is no widely adopted taxonomy that jointly covers TFA types, preventive measures, detection mechanisms, and support services. This makes it difficult to reason about coverage and gaps in current responses, both in the research literature and in the web resources that people encounter when seeking help. It also limits the creation of reusable datasets for vulnerable populations and hinders replicable, data driven analyses of how institutional web content contributes to or mitigates harm.

To address this gap, we first ask how prior work conceptualises the life cycle of TFA, from prevention before abuse occurs to detection while it is ongoing, and subsequently, support after harm has taken place. We then examine how this landscape is reflected in web-based resources provided by institutions, focusing on public and social service websites that people are likely to encounter when they search for information or assistance. Our study is guided by the following research questions:

\textbf{RQ1}: What are the different types of TFA, and which aspects, such as detection, prevention, and support, have been addressed in the literature?

\textbf{RQ2}: What facilities or support services are currently available through the government, non-government organisations (NGOs), and related agencies on the web, and how are these resources framed emotionally and presented in terms of readability and accessibility?

\textbf{RQ3}: What are the existing shortcomings in current responses to TFA, and what strategies can be implemented to strengthen future prevention and support efforts?

To answer these questions, we conduct a structured review of 38 peer-reviewed papers and practitioner reports on TFA and related harms~\cite{bellini2023paying, o2023short, wang2025breaking, liggett2025suicidal, bellini2023digital, dwyer2025friend, flynn2021technology, flynn2022technology, chen2022trauma, brown2021technology, scott2023trauma,freed2019my,stephenson2023abuse,abhinaya2024enabling,stephenson2023s,henry2020technology,sheikh2024technology,obada2022sok,wei2024understanding,flynn2024workplace,qin2024did,heinrich2024please,turk2024stop,turk2023can,heinrich2022airguard,ceccio2023sneaky,pace2023every,mangeard2023no,havron2019clinical,almansoori2024web,wei2022anti,coduto2024delete,stephenson2024sharenting,bellini2024abusive,mckay2021standing,gupta2024really,chowdhury2024safe}. From this extensive review, we derive four interconnected taxonomies that capture (1) TFA types, (2) prevention strategies before abuse, (3) detection mechanisms during abuse, and (4) support services after abuse. Together, this TFA taxonomy defines a common set of categories that can be used to (i) benchmark how different web ecosystems cover the TFA life cycle in the resources they offer to victim survivors, (ii) monitor institutional web responses over time by reapplying the taxonomy to new crawls or policy updates, (iii) build labelled datasets for training and evaluating classification or retrieval systems that surface TFA-related resources, and (iv) provide a concrete checklist for designing or auditing public-facing websites and policies. It therefore serves both as an analytic lens for this study and as a reusable scaffold for future, comparable, data-driven work on TFA-related web resources.

As an Australian case study of how this taxonomy can be applied to real-world web infrastructures, we investigate the current state of web-based resources for people affected by TFA by analysing institutional domains registered in Australia. We crawl 306 government, non-government, and relevant commercial domains and collect 52\,605 unique web pages that contain content related to domestic and family violence, online safety, or digital abuse. Using a hierarchical zero-shot topic modelling pipeline based on our TFA taxonomy, we perform multi-label classification of each page into abuse types and corresponding prevention, detection, and support categories. Beyond taxonomy-aligned classification, we automatically characterise the dominant emotion and readability of each page using a RoBERTa-based emotion classifier and a suite of readability metrics, which allows us to study not only what information is present but also how it is framed and how accessible it is to different audiences. Here, we use the term ``audit'' in the sense common to web research, namely a systematic, data-driven assessment of the information exposed through institutional web pages, evaluated against a structured taxonomy of expected content. Our analysis reveals the following: 

\begin{enumerate}
    \item Web resources are dominated by three abuse types i.e., \textit{Comments Abuse} (30\% of the corpus), \textit{Harassment} (26\%), and \textit{Sexual Abuse} (14\%), which together account for roughly two thirds of all pages. By contrast, \textit{Economic Abuse}, \textit{Privacy Violation}, and \textit{Controlling Behaviours} each appear on fewer than 1\% of pages.
    
    \item In the prevention and detection stages, institutional pages primarily foreground legal and clinical responses. For prevention, \textit{Legal Measures} account for 49\% of pages and \textit{Family Safeguarding} for 15\%, while \textit{Technology Based Safeguards}, \textit{Public Awareness and Education}, and \textit{Digital Community Networks} together cover less than 1\%. For detection, \textit{Clinical Detection} (49\%) and \textit{Incidental Discovery} (21\%) dominate, whereas \textit{Institutional and Community Reporting} and \textit{Device Detection} are effectively absent.
    
    \item Support-related content is concentrated in generic digital help and technical safety advice. \textit{Digital Support Infrastructure} accounts for 62\% of support pages and \textit{Technology Security Support} for 8\%, while \textit{Institutional Support}, \textit{Psychological Support}, \textit{Informal Support Networks}, and \textit{Legal Support} each cover fewer than 1\% of pages, with legal support never appearing as a distinct category.
    
    \item Emotion and readability profiling shows that institutional content is often emotionally neutral but linguistically demanding, and that underrepresented harms tend to be framed with more intense negative affect. Core categories such as \textit{Digital Support Infrastructure}, \textit{Legal Measures}, and \textit{Harassment} are predominantly neutral in tone (around 40\% to 50\% of pages per category) and are written at reading levels associated with late secondary or early tertiary education. In contrast, rare categories such as \textit{Economic Abuse} and \textit{Privacy Violation} are dominated by fear (up to 90.5\% of pages for \textit{Economic Abuse}) and have lower accessibility scores, making them both harder to locate and harder to comprehend.
\end{enumerate}

These findings highlight substantial gaps between the harms documented in the literature and the help that is visible to people who rely on institutional websites for information and support around TFA. They also illustrate how web scale audits grounded in a shared TFA taxonomy can make institutional responses more transparent, enable collaboration between data owners and problem owners, and inform the design of web resources that better promote safety, equity, and social impact for those affected by TFA. The remainder of the paper is organised as follows. Section~\ref{sec:taxonomy} introduces the TFA taxonomy derived from the literature. Section~\ref{sec:case-study} presents our Australian web case study, including data collection, taxonomy-aligned classification, and emotion and readability assessment. We conclude in Section~\ref{sec:conclusion}.

%% file: sections/02-tfa-taxonomy.tex
\section{TFA Taxonomy Creation}
\label{sec:taxonomy}

In this section, we explain how we derived the TFA taxonomy based on prior work. In line with the lifecycle perspective outlined in Section~\ref{sec:intro}, the taxonomy distinguishes between (1) abuse types, (2) prevention strategies before or between incidents, (3) detection mechanisms while abuse is ongoing, and (4) support measures after harm has occurred. We later use this shared label space to analyse institutional web content at scale.

\subsection{Literature Search}

Given the rapid evolution of TFA, we focused on recent work. We queried Google Scholar and major digital libraries using combinations of terms such as ``technology facilitated abuse'', ``tech abuse'', ``intimate partner surveillance'', ``image based abuse'', and ``online harassment'', targeting leading venues in security, privacy, and human computer interaction (for example CHI, SOUPS, USENIX Security, IEEE Security \& Privacy) and relevant journals. We restricted our corpus to papers published after 2020, with one earlier foundational review~\cite{henry2020technology}. After screening titles and abstracts, we retained 38 papers and practitioner reports that provided empirical data or conceptual frameworks on TFA. Based on the papers, we define four dimensions that mirror the abuse life cycle: (1) abuse types, (2) prevention strategies, (3) detection mechanisms, and (4) support measures. Through iterative coding and constant comparison, we grouped similar concepts into higher-level categories, which form the four taxonomies.

\subsection{TFA Types Taxonomy}
\label{subsec:tfa-types}

\begin{figure*}[ht!]
  \centering
  \includegraphics[width=0.93\linewidth]{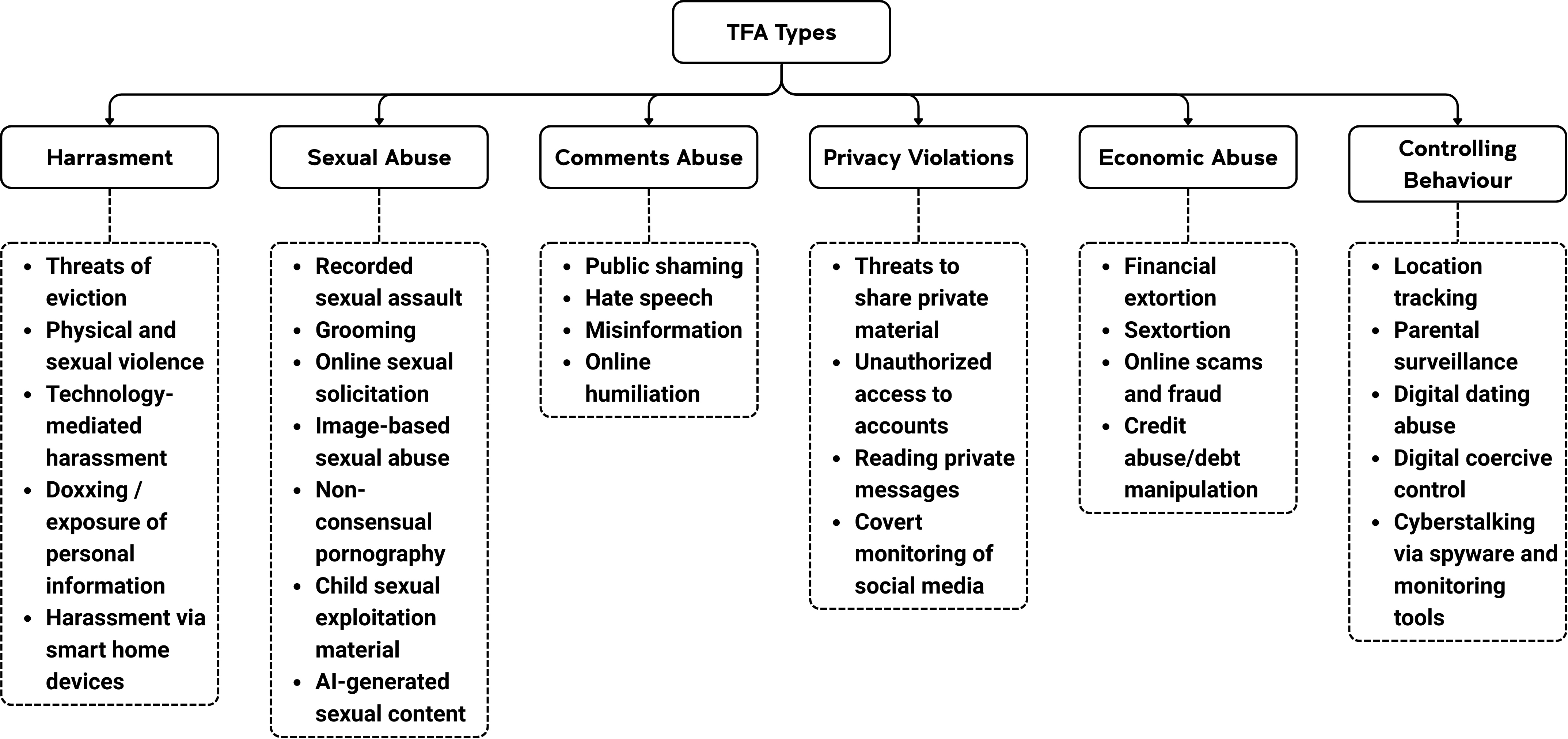}
  \caption{Taxonomy of technology facilitated abuse types.}
  \Description{A taxonomy diagram illustrating various types of technology facilitated abuse, such as harassment, sexual abuse, economic abuse, privacy violations, comments abuse, and controlling behaviours.}
  \label{fig:abuse-types}
\end{figure*}

Building on typologies of digital dating abuse, cyberstalking, image-based abuse, and online sexual harassment~\cite{henry2020technology}, we group TFA into six main categories (Figure~\ref{fig:abuse-types}) and refine each with subcategories grounded in prior work: 

\begin{itemize}
  \item \textbf{Harassment} captures threatening or repetitive abuse via digital channels, informed by studies of online harassment and workplace digital sexual harassment~\cite{henry2020technology,flynn2024workplace,sheikh2024technology}. Subcategories include direct threats of eviction or physical and sexual violence, technology-mediated harassment via email, messaging, and mobile apps, doxxing or exposure of personal information, and harassment through smart home and VR systems~\cite{stephenson2023s,abhinaya2024enabling}.
  \item \textbf{Sexual abuse} covers coercive or non-consensual sexual behaviours facilitated by technology, including image-based sexual abuse, non-consensual pornography, and financially motivated sextortion~\cite{obada2022sok,wei2024understanding,liggett2024suicidal,o2023short,qin2024did}. Subcategories include recorded sexual assault, grooming, online sexual solicitation, image-based abuse and non-consensual pornography, child sexual exploitation material, and AI-generated sexual content~\cite{stephenson2024sharenting}.
  \item \textbf{Comments abuse} includes public shaming, hate speech, misinformation, and online humiliation, drawing on studies of digital harassment in workplaces and social platforms~\cite{flynn2024workplace}.
  \item \textbf{Privacy violations} involve unauthorized access, covert surveillance, and exposure of private information~\cite{havron2019clinical,heinrich2024please,mangeard2023no,chen2022trauma}. Subcategories include threats to share private or financial material, posting financial information online, unauthorized access to accounts, reading private messages, and covert monitoring of social media~\cite{stephenson2024sharenting}.
  \item \textbf{Economic abuse} draws on research on technology mediated financial coercion and exclusion~\cite{bellini2023paying,bellini2023digital,dwyer2025friend}. Subcategories include financial extortion and sextortion~\cite{o2023short,wang2025breaking,liggett2024suicidal}, online scams and fraud, credit abuse or debt manipulation in the victim's name, and reputation attacks that cause economic harm.
  \item \textbf{Controlling behaviours} include location tracking, parental surveillance, digital dating abuse, coercive control, and cyberstalking via spyware and monitoring tools~\cite{brown2021technology,gupta2024really,heinrich2022airguard,heinrich2024please,turk2024stop,turk2023can,freed2019my,stephenson2023abuse,wei2022anti,stephenson2024sharenting}.
\end{itemize}

This taxonomy provides the label space for our abuse type classification, allowing us to distinguish, for example, pages focused on sextortion from those focused on parental tracking.

\begin{figure*}[ht!]
  \centering
  \includegraphics[width=0.93\linewidth]{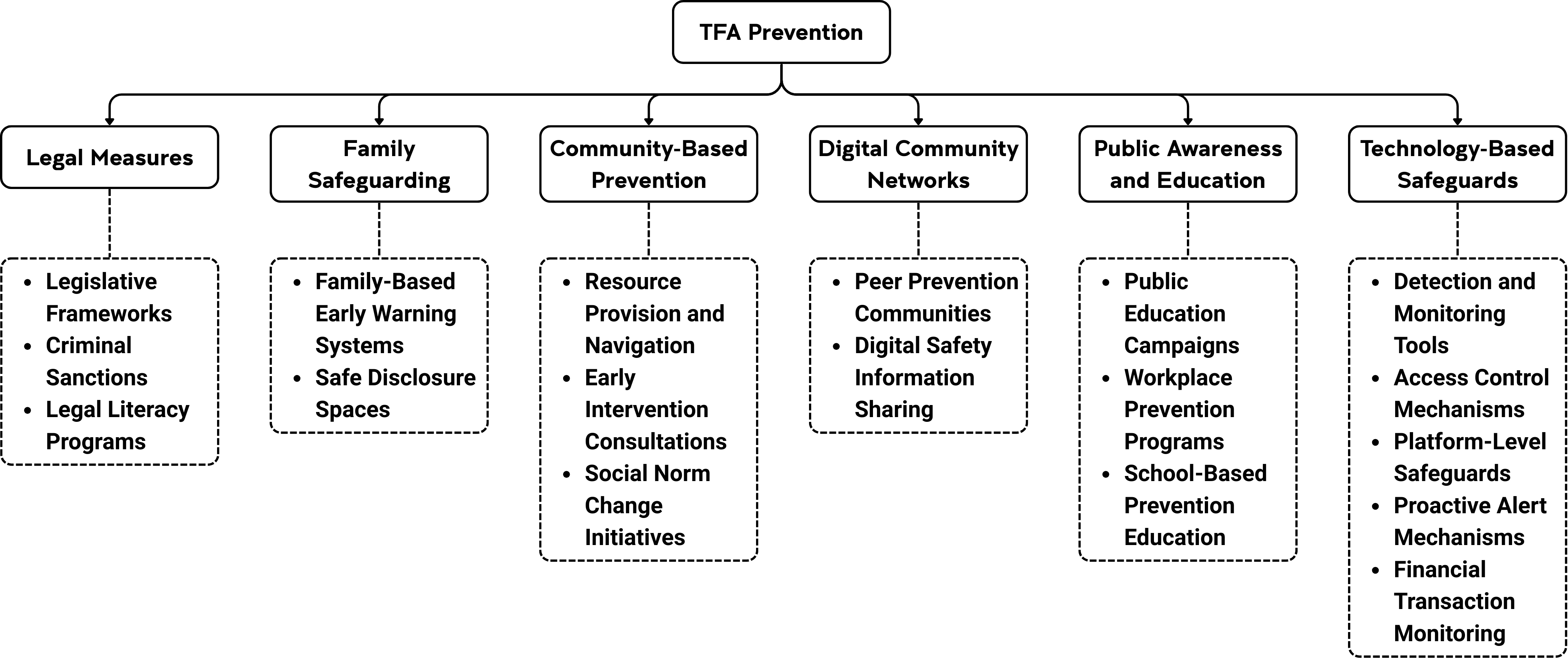}
  \caption{Taxonomy of technology facilitated abuse prevention strategies.}
  \Description{A taxonomy diagram illustrating prevention types, including legal frameworks, family safeguarding, community based prevention, digital networks, education campaigns, and technology based safeguards.}
  \label{fig:abuse-prevention}
\end{figure*}

\subsection{TFA Prevention Strategies Taxonomy}
\label{subsec:tfa-prevention}

Prevention work, as depicted in Figure~\ref{fig:abuse-prevention} spans legal, social, and technical interventions that aim to reduce the risk of TFA before or between incidents:

\begin{itemize}
  \item \textbf{Legal Measures} include criminalisation of image-based abuse, cyberstalking, and coercive control, civil penalties, and legal literacy programs that inform people about rights and protections~\cite{henry2020technology,gupta2024really,chowdhury2024safe,CalSharentingLaw2024,IllSharentingLaw2023}.
  
  \item \textbf{Family Safeguarding} covers early warning and safe disclosure mechanisms for children, older adults, disabled people, and other vulnerable groups~\cite{sheikh2024technology,gupta2024really,stephenson2023s}. Subcategories include observing changes in emotional or digital behaviour and balancing protective monitoring with risks of over-surveillance~\cite{wei2022anti,stephenson2024sharenting}.
  
  \item \textbf{Community-based Prevention} includes NGO led resource provision and navigation, early intervention consultations, and efforts to shift social norms around harassment, victim blaming, and gender inequality~\cite{henry2020technology,wei2024understanding,sheikh2024technology,gupta2024really}.
  
  \item \textbf{Digital Community Networks} provide prevention-oriented peer support and early warning information through online platforms that share red flags and digital safety practices~\cite{sheikh2024technology,gupta2024really}.
  
  \item \textbf{Public Awareness and Education} include public campaigns, workplace policies and training~\cite{flynn2024workplace}, and school-based curricula on online safety, healthy relationships, consent, and digital ethics~\cite{henry2020technology}.
  
  \item \textbf{Technology-based Safeguards} include anti stalking and tracker detection apps~\cite{heinrich2022airguard,heinrich2024please}, safer access control and notification design for smart devices~\cite{stephenson2023abuse}, and financial sector tools to detect abuse related transactions~\cite{bellini2023paying,bellini2023digital}.
\end{itemize}

In our web analysis, these labels indicate whether a page presents TFA only as a problem or also points to concrete preventive measures, and of what kind.

\subsection{TFA Detection Mechanisms Taxonomy}
\label{subsec:tfa-detection}

\begin{figure*}[ht!]
  \centering
  \includegraphics[width=0.93\linewidth]{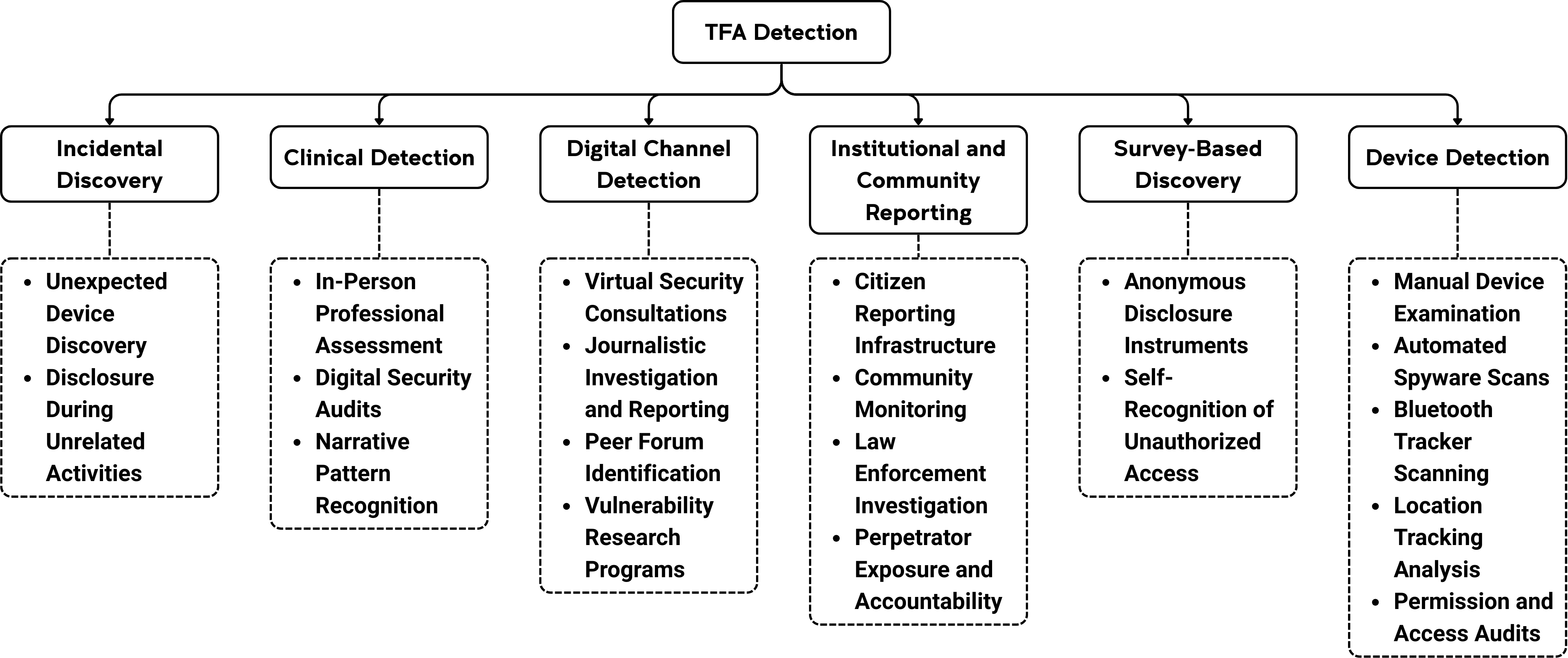}
  \caption{Taxonomy of technology-facilitated abuse detection mechanisms.}
  \Description{A taxonomy diagram illustrating detection categories, including incidental discovery, clinical consultations, online and media-based detection, civic reporting systems, survey-prompted awareness, and device-level detection.}
  \label{fig:abuse-detection}

\end{figure*}

The transition from hidden or normalised behaviour to recognised abuse often depends on how TFA is detected. Prior work points to several distinct mechanisms, as presented in Figure~\ref{fig:abuse-detection}:

\begin{itemize}
  \item \textbf{Incidental Discovery} covers cases where abuse is uncovered unintentionally, such as the unexpected discovery of trackers, hidden cameras, or monitoring tools, or disclosure of tech abuse during unrelated therapy, legal, or support encounters~\cite{stephenson2023s}.
  
  \item \textbf{Clinical Detection} involves in-person professional assessments and digital security audits in which therapists, lawyers, social workers, or security specialists combine survivor narratives with structured device and account checks and recognise patterns of coercive control or surveillance~\cite{freed2019my,havron2019clinical,chen2022trauma}.
  
  \item \textbf{Digital Channel Detection} includes remote security clinics~\cite{chen2022trauma}, peer forums where survivors recognise patterns in shared narratives~\cite{wei2024understanding,wang2025breaking}, and vulnerability research programs that expose abuse vectors in platforms and services~\cite{bellini2023digital}.
  
  \item \textbf{Institutional \& Community Reporting} covers citizen reporting infrastructures such as online forms and hotlines, community monitoring and moderation, and law enforcement or regulatory investigations that respond to reports of TFA~\cite{henry2020technology,wang2025breaking,flynn2021technology,bellini2024abusive,sheikh2024technology}.
  
  \item \textbf{Survey-Based Discovery} reflects the role of anonymous questionnaires and interviews in helping participants identify their experiences as abuse and in documenting patterns of monitoring, threats, and coercion~\cite{flynn2021technology,flynn2022technology,brown2021technology,stephenson2023s}.
  
  \item \textbf{Device Detection} includes manual examination of devices and accounts, automated spyware scans~\cite{freed2019my}, Bluetooth tracker scanning and forensic reconstruction of stalking histories~\cite{heinrich2024please,pace2023every}, classifier-based detection of suspicious online behaviour~\cite{almansoori2024web}, and audits that reveal privacy and security flaws that can be abused~\cite{ceccio2023sneaky,bellini2023digital}.
\end{itemize}

Together with the prevention taxonomy, these detection labels let us assess whether institutional pages point towards concrete ways of identifying TFA, such as device scans or reporting channels, or stay at a purely conceptual level.

\begin{figure*}[ht!]
  \centering
  \includegraphics[width=0.93\linewidth]{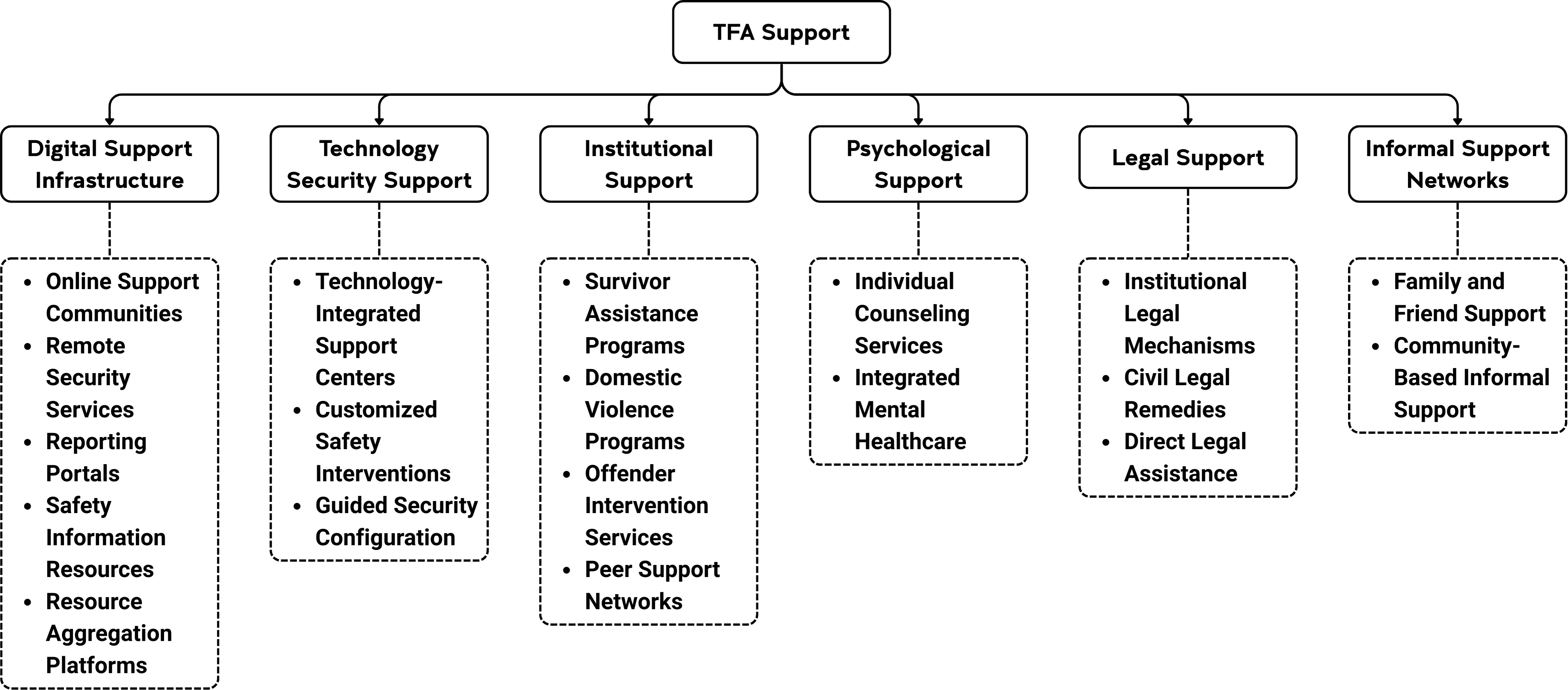}
  \caption{Taxonomy of technology-facilitated abuse support measures.}
  \Description{A taxonomy diagram illustrating support types, including digital support infrastructure, digital safety assistance, programmatic support services, mental health support, legal aid and advocacy, and personal network support.}
  \label{fig:abuse-support}
\end{figure*}

\subsection{TFA Support Measures Taxonomy}
\label{subsec:tfa-support}

Once abuse is recognised, survivors may seek ongoing support across technical, social, legal, and psychological domains. We present the taxonomy in Figure~\ref{fig:abuse-support}. Prior work describes a mix of formal and informal pathways, which we group into six categories:

\begin{itemize}
  \item \textbf{Digital Support Infrastructure} bundles online support communities~\cite{sheikh2024technology,gupta2024really}, remote security services and guidance~\cite{chen2022trauma,almansoori2024web}, official reporting portals~\cite{henry2020technology}, and curated safety information resources.
  
  \item \textbf{Technology Security Support} covers specialised help for device security, account protection, and safety configuration, including technology-integrated support centres, customised safety interventions, and guided configuration help for privacy settings and authentication~\cite{havron2019clinical,gupta2024really}.
  
  \item \textbf{Institutional Support} includes structured survivor assistance programs, domestic and family violence services, and offender intervention programs that combine mediation, behaviour change, and accountability~\cite{bellini2024abusive,flynn2021technology}.
  
  \item \textbf{Psychological Support} includes individual counselling and trauma-focused therapy for survivors of sextortion, image-based abuse, and other TFA~\cite{wang2025breaking,liggett2024suicidal}, and integrated mental health care in redress and victim support schemes.
  \item \textbf{Legal Support} spans institutional legal mechanisms such as criminal justice and compensation, civil legal remedies including protection orders and damages claims, and direct legal assistance and advocacy by NGOs~\cite{henry2020technology}.
  
  \item \textbf{Informal Support Networks} reflects assistance from friends, family, and community organisations, which often act as first responders and intermediaries when survivors seek help~\cite{flynn2022technology,sheikh2024technology,gupta2024really}.
\end{itemize}
These four taxonomies operationalise the TFA lifecycle introduced in Section~\ref{sec:intro} along four dimensions: what kinds of abuse occur, how they might be prevented, how they are detected, and what support is offered afterwards. In the remainder of the paper, we use this shared label space to classify Australian institutional websites to quantify where online resources align with, or fall short of, the TFA prevention, detection, and support needs.

%% file: sections/03-case-study.tex
\section{Case Study of Australian Domains}\label{sec:case-study}
Australia offers a uniquely important case study for TFA because the country is simultaneously experiencing some of the highest recorded rates of domestic, family, and sexual violence among OECD nations, and is also home to ambitious national initiatives aimed at prevention, digital safety, and survivor support\footnote{\url{https://www.esafety.gov.au/women/reduce-technology-facilitated-abuse}; Accessed on 28 November 2025}. Recent surveys show that 1 in 4 Australian women and 1 in 8 men have experienced intimate-partner violence, and an estimated 65-70\% of domestic-violence survivors report technology-enabled harassment, monitoring, or threats, placing TFA at the centre of national policy concern~\cite{AIHW2023IPV, BOSCAR2024DV, ANROWS2022TFA}. The Australian Government has invested more than \$2.3 billion under the National Plan to End Violence Against Women and Children (2022-2032), including targeted programs addressing online harms, safety by design, and tech-abuse prevention in partnership with eSafety, state governments, and frontline services~\cite{NationalPlan2022, AustralianGov2024Investment}. Compared to many countries, Australia also maintains a strong institutional infrastructure, some being the eSafety Commissioner, national domestic-violence hotlines, and state-level women’s safety initiatives, making it an ideal environment to evaluate how public institutions communicate risks and support pathways around TFA.

\subsection{Data Collection}

\paragraph{Domain selection.}
We first identified Australian web domains likely to contain TFA-related content using advanced Google search operators (for example, \texttt{site:.au} combined with 32 TFA-specific keywords derived from our literature review). This yielded 2,212 candidate domains referencing phenomena such as ``image-based abuse'', ``online harassment'', and ``technology-facilitated coercive control''. We manually screened all domains and retained 306 that provided substantive information or support related to domestic and family violence, online safety, or digital abuse. General media sites (e.g., \texttt{9news.com.au}), where TFA keywords occurred only sporadically in isolated articles, were excluded. By contrast, domains whose primary purpose is related to child protection, safety, or advocacy (such as \texttt{actparents.org.au}) were retained.

\paragraph{Crawling and content extraction.}
For each domain, we deployed a custom web crawler based on depth-first search (DFS) with a maximum traversal depth of four. Crawling began from a domain-specific entry URL and followed internal links either across the full domain (for small and medium sites) or within chosen subdirectories (for large sites). Newly discovered internal URLs were added to a stack until the depth limit was reached. Unsupported MIME types, such as spreadsheets, compressed archives, and multimedia, were discarded. HTML pages were rendered using Playwright with a headless Chromium backend to ensure complete DOM extraction, and the resulting HTML source was stored in our database. Extremely large domains that would require more than one hour of crawling were handled manually by identifying TFA-relevant entry points and restricting traversal to URLs sharing those prefixes.

\paragraph{Preprocessing.}
The raw crawl produced 56,672 URLs across 306 domains. We applied a multi-stage preprocessing pipeline to extract visible textual content, removing boilerplate such as navigation menus, scripts, and style blocks. We further filtered pages with empty or near-empty text, pages below a minimum length threshold, and pages failing an English-language check. The final dataset comprises 52,605 cleaned pages (Table~\ref{tab:data-stat}).

\subsection{Taxonomy-Aligned Classification}

To align the corpus with the four taxonomy dimensions introduced in Section~\ref{sec:taxonomy}, we perform hierarchical multi-label text classification. Each cleaned page is mapped to one or more primary categories in each taxonomy and, where supported by sufficient evidence, to corresponding subcategories. All label predictions are constrained strictly to the predefined taxonomy structures.

\paragraph{Primary category classification.}
For each taxonomy (abuse, prevention, detection, support), we classify web pages at the level of primary categories. We embed each web page using all-MiniLM-L6-v2\footnote{\url{https://huggingface.co/sentence-transformers/all-MiniLM-L6-v2}}, a SentenceTransformer model~\cite{reimers-2019-sentence-bert}, and apply zero-shot BERTopic~\cite{grootendorst2022bertopicneuraltopicmodeling} with the primary taxonomy categories as candidate labels. The model uses (i) all-MiniLM-L6-v2 for embeddings, (ii) deberta-v3-large-zeroshot-v2.0\footnote{\url{https://huggingface.co/MoritzLaurer/deberta-v3-large-zeroshot-v2.0}} as a zero-shot similarity model, and (iii) a similarity threshold of $0.5$. Categories meeting or exceeding this threshold are retained, yielding a multi-label assignment.

\paragraph{Subcategory classification.}
For each primary category, we select all web pages assigned that label and apply a second zero-shot BERTopic model restricted to the subcategories of that primary category. This produces a refined, category-specific multi-label assignment. To avoid unstable models on small subsets, subcategory classification is performed only when a primary category contains at least 15 web pages; otherwise the taxonomy assignment remains at the primary level. A similarity threshold of $0.5$ is again applied.

\subsection{Emotion and Readability Assessment}

To understand not only \textit{what} institutions say about TFA but also \textit{how} they communicate it, we analyse each web page along two complementary axes: affective framing and textual accessibility. Emotion profiles indicate whether institutional narratives emphasise fear, neutrality, reassurance, or other stances, while readability metrics capture how easy it is for survivors, practitioners, and the general public to engage with the information.

\paragraph{Emotion classification.}
We apply a RoBERTa-based~\cite{liu2019robertarobustlyoptimizedbert} English emotion classifier\footnote{\url{https://huggingface.co/j-hartmann/emotion-english-distilroberta-base}} that outputs seven mutually exclusive emotions: anger, disgust, fear, joy, neutral, sadness, and surprise. For each web page we retain the dominant emotion (highest-scoring class). At the taxonomy level, we compute the proportion of web pages whose dominant emotion corresponds to each label, yielding the emotion distributions.

\paragraph{Readability and accessibility.}
We compute readability indicators using \texttt{textstat}\footnote{\url{https://docs.textstat.org/}}, including Flesch Reading Ease (higher indicates easier text) and the Automated Readability Index (higher indicates more complex text). We also define a composite accessibility score (0-100) that penalises long sentences, long paragraphs, and high grade-level scores. Category-level readability scores are obtained by averaging over all web pages assigned to each taxonomy category.

\input{tables/data-stats}

\subsection{Analysis of Taxonomy Coverage, Emotion, and Accessibility}

Table~\ref{tab:tfa-analysis} summarises how the 52,605 web pages in our corpus are distributed across the four taxonomy dimensions, and how these categories differ in affective tone and linguistic accessibility. Below, we interpret these results in relation to our three research questions.

\input{tables/main-results}

\subsubsection{RQ1: Coverage of TFA types, detection, prevention, and support}
The taxonomy-aligned classification operationalises the literature-derived conceptual model of TFA (Section~\ref{sec:taxonomy}) at scale. Across abuse types, institutional web content overwhelmingly foregrounds \textit{Comments Abuse} (30\%) and \textit{Harassment} (26\%). \textit{Sexual Abuse} appears in 14\% of the corpus, but other categories identified in the TFA literature are underrepresented in the Australian domains. These include \textit{Privacy Violation}, \textit{Economic Abuse}, and \textit{Controlling Behaviours}, each appearing in fewer than 1\% of pages. Their emotion profiles further signal their marginalisation: pages on \textit{Privacy Violation} are dominated by \textit{Fear} (38.1\%) and \textit{Anger} (33.6\%), while \textit{Economic Abuse} is overwhelmingly classified as \textit{Fear} (90.5\%). A similar pattern arises for detection, prevention, and support. In detection, institutional content is dominated by \textit{Clinical Detection} (49\%) and \textit{Incidental Discovery} (21\%), whereas categories representing systematic detection infrastructures such as \textit{Institutional and Community Reporting} and \textit{Device Detection} do not appear in the corpus at all. Within prevention, nearly half of all relevant pages concern \textit{Legal Measures} (49\%), followed by \textit{Family Safeguarding} (15\%). As shown in Table~\ref{tab:tfa-cooccurrence} (Appendix~\ref{sec:cooccurence}), the most common subcategories within these types reinforce this focus on visible, public-facing harms. For instance, the dominant \textit{Comments Abuse} subcategory is \textit{Public Shaming}, \textit{Harassment} is often instantiated as \textit{Threats of Eviction}, and \textit{Sexual Abuse} is frequently framed as \textit{Recorded Sexual Assault}. In contrast, there are far fewer pages that describe covert surveillance, financial coercion, or long-term controlling behaviours in detail, despite their prominence in the research taxonomy. The most frequent detection subcategories, \textit{In Person Professional Assessment} and \textit{Unexpected Device Discovery}, show that TFA is often framed as something that comes to light during face-to-face consultations or by chance, rather than through dedicated tools or reporting systems. Prevention measures are instantiated as \textit{Criminal Sanctions} and \textit{Family-Based Early Detection}. Categories emphasised in research as forward-looking and technology-centric, such as \textit{Technology Based Safeguards} and \textit{Public Awareness and Education}, together account for less than 1\% of pages. Support content is similarly skewed: \textit{Digital Support Infrastructure} accounts for 62\% of support-related pages, and its most common subcategory, \textit{Online Support Communities}, appears on 25,912 (out of 52,605) pages. By comparison, \textit{Technology Security Support}, \textit{Institutional Support}, and \textit{Psychological Support} have far fewer pages and smaller subcategory counts.



\subsubsection{RQ2: Web-facing facilities and support services}
For survivors, practitioners, or the public accessing institutional websites, the most visible and accessible facilities are those tied to \textit{Digital Support Infrastructure} (62\%) and \textit{Technology Security Support} (8\%). These categories also display largely neutral emotion profiles (45.6\% and 40.4\% neutral respectively) and mid-range accessibility scores (59.70 and 58.13), which is consistent with informational self-help content aimed at broad audiences. The distribution of support-related categories indicates that institutional websites overwhelmingly prioritise informational guidance over actionable service pathways. Almost two-thirds of all support pages fall under \textit{Digital Support Infrastructure}, whereas categories directly associated with service access such as \textit{Institutional Support}, \textit{Psychological Support}, \textit{Informal Support Networks}, and \textit{Legal Support} collectively account for under 2\% of pages, with \textit{Legal Support} entirely absent. As shown in Table~\ref{tab:support-examples} (Appendix~\ref{sec:examples}), sites associated with \textit{Digital Support Infrastructure}, such as \texttt{southsafe.org.au}, \texttt{fcfcoa.gov.au}, and \texttt{ndiscommission.gov} \texttt{.au}, primarily offer static information hubs, downloadable resources, and broad safety guidance. Content mapped to \textit{Technology Security Support} on domains such as \texttt{gcyp.sa.gov.au} or \texttt{legalaid.vic.gov} \texttt{.au} focuses on high level digital safety advice, device configuration, and account protection, but often does not provide direct mechanisms for live digital security assistance. By contrast, \textit{Institutional Support} and \textit{Psychological Support} tend to appear as brief descriptions of programs or counselling services that require additional navigation or offline steps to access. Mentions of \textit{Informal Network Support} are particularly sparse, appearing only on a small number of pages.


\subsubsection{RQ3: Gaps and opportunities for strengthening responses}
The combination of coverage, emotion, readability, and subcategory patterns reveals systematic gaps that point to concrete avenues for strengthening TFA responses. First, there is a pronounced gap between the harms emphasised in the literature and those visible online. Categories such as \textit{Privacy Violation} and \textit{Economic Abuse} are nearly absent (less than 1\%) and are emotionally dominated by \textit{Fear} (38.1\% and 90.5\% respectively). Their low volume and high emotional intensity risk making them both difficult to locate and cognitively burdensome to process. Expanding coverage and presenting clearer, stepwise guidance in these areas would better align institutional communication with empirical TFA research. Second, prevention and detection content remains largely reactive rather than proactive. The absence of \textit{Technology Based Safeguards}, \textit{Institutional and Community Reporting}, and \textit{Device Detection} suggests that institutional sites undercommunicate actionable, technology-driven countermeasures, even though these feature prominently in technical and policy literatures on anti stalking, tracker detection, and safety by design. Incorporating descriptions of detection tools, reporting workflows, and platform-level protections could materially strengthen web-facing prevention and detection. Third, readability metrics indicate that many categories fall within late secondary or early tertiary reading levels. Automated Readability Index (ARI) values for central categories often lie between 18 and 22, with accessibility scores in the range of 50 to 60. For instance, the \textit{Harassment} category reflects an ARI of 20.27 and an accessibility score of 59.16, while \textit{Comments Abuse} shows ARI 21.48 with a score of 59.46. According to recent data from the Australian Bureau of Statistics\footnote{\url{https://www.abs.gov.au/statistics/people/education}}, approximately 73\% of Australians aged 25 to 34 hold a Certificate III qualification or higher, meaning that nearly one in four adults may have lower formal education. Content written at tertiary reading levels may therefore pose undue difficulty for many at risk users. Notably, underrepresented categories such as \textit{Economic Abuse} and \textit{Privacy Violation}, despite their emotional intensity, receive less than 1\% of pages and often use complex language and dense structure. By contrast, \textit{Informal Support Networks}, although it occupies less than 1\% of the corpus, achieves a higher Flesch Reading Ease score (55.48) and accessibility score (66.50), illustrating that more accessible content is possible but rare. Co design with community partners may therefore help institutions produce guidance that is both clearer and more supportive~\cite{flynn2021technology,gupta2024really}.

\textit{\textbf{Takeaway:} Our analysis shows that Australian institutional web content is strongest where it provides legal framing, high-level information hubs, and digital safety guidance, but weaker in conveying covert harms, technological prevention, systematic detection, and multidimensional support pathways.} 



%% file: tables/data-stats.tex
\begin{table}[t!]
\centering
\begin{adjustbox}{center,width=\linewidth}
\begin{tabular}{l|c|c}
\hline
\textbf{Top 10 Domains} & $\bm{\mu_{\text{word}}}$ & \textbf{Count} \\
\hline
southsafe.org.au & 368.8 & 3297 \\
fcfcoa.gov.au & 511.9 & 2129 \\
ndiscommission.gov.au & 264.6 & 1864 \\
legalaid.wa.gov.au & 314.8 & 1717 \\
gcyp.sa.gov.au & 269.5 & 1589 \\
professionals.childhood.org.au & 455.2 & 1428 \\
lwb.org.au & 532.1 & 1256 \\
criminal-lawyers.com.au & 668.9 & 1207 \\
legalaid.vic.gov.au & 610.1 & 1193 \\
headtohealth.gov.au & 150.3 & 1191 \\
\hline
\textbf{Total} & \textbf{397.8} & \textbf{52605} \\
\hline
\end{tabular}
\end{adjustbox}
\caption{Average number of words per page $\bm{\mu_{\text{word}}}$ and total page counts for the ten largest domains in the corpus, together with aggregate statistics for the full dataset.}
\vspace{-8mm}
\label{tab:data-stat}
\end{table}

%% file: tables/main-results.tex
\begin{table*}[ht!]
\centering
\begin{adjustbox}{center,width=0.88\linewidth}
\begin{tabular}{l|c|ccccccc|ccc}
\hline
\multirow{2}{*}{\textbf{Category}} &
\multirow{2}{*}{\textbf{Corpus (\%) }} &
\multicolumn{7}{c|}{\textbf{Emotion (\%) }} &
\multicolumn{3}{c}{\textbf{Readability}} \\
\cline{3-12}
& & Neutral & Joy & Fear & Sadness & Anger & Surprise & Disgust & Flesch ($\uparrow$) & ARI ($\downarrow$) & Access. ($\uparrow$) \\
\hline
\multicolumn{12}{c}{\textbf{TFA Types}} \\
\hline
Harassment                 & 26  & \textbf{41.5} & 21.4 & 13.4 & 12.2 & 6.9 & 3.9 & 0.7 & 25.07 & 20.27 & 59.16 \\
Sexual Abuse               & 14  &\textbf{40.6} & 19.4 & 11.7 & 14.4 & 9.2 & 3.8 & 0.9 & 30.49 & 18.01 & 59.18 \\
Comments Abuse             & 30  & \textbf{48.3} & 18.6 & 11.3 & 11.5 & 5.4 & 4.3 & 0.6 & 24.00 & 21.48 & 59.46 \\
Privacy Violation          & <1 & 17.2 & 2.8 & \textbf{38.1} & 6.3 & 33.6 & 0.4 & 1.8 & 31.58 & 16.56 & 42.50 \\
Economic Abuse             & <1 & 0.3 & 0.2 & \textbf{90.5} & 0.3 & 8.6 & 0.1 & 0.1 & 31.10 & \textbf{11.70} & 65.00 \\
Controlling Behaviours     & <1 & \textbf{81.5} & 3.5 & 4.0 & 2.1 & 4.2 & 2.2 & 2.5 & \textbf{31.95} & 14.97 & \textbf{69.62} \\
\hline
\multicolumn{12}{c}{\textbf{TFA Prevention}} \\
\hline
Legal Measures                 & 49 & \textbf{44.6} & 18.1 & 11.4 & 13.6 & 7.6 & 4.1 & 0.7 & 27.08 & 19.92 & 59.55 \\
Family Safeguarding            & 15 & \textbf{47.0} & 18.9 & 12.5 & 11.2 & 5.7 & 3.9 & 0.8 & 19.35 & 22.41 & 57.57 \\
Community-Based Prevention     & 6  & \textbf{41.5} & 24.6 & 13.3 & 12.0 & 4.3 & 3.9 & 0.5 & 26.22 & 20.64 & 60.46 \\
Digital Community Networks     & 0 & -- & -- & -- & -- & -- & -- & -- & -- & -- & -- \\
Public Awareness and Education & <1  & \textbf{53.9} & 11.9 & 25.8 & 1.8 & 3.9 & 2.5 & 0.3 & -8.68 & 29.03 & 42.06 \\
Technology-Based Safeguards    & <1  & 23.0 & \textbf{50.5} & 2.9 & 2.7 & 17.7 & 3.1 & 0.2 & \textbf{38.98} & \textbf{13.87} & \textbf{75.00} \\
\hline
\multicolumn{12}{c}{\textbf{TFA Detection}} \\
\hline
Clinical Detection             & 49 & \textbf{51.3} & 20.7 & 8.9 & 11.8 & 4.0 & 2.8 & 0.6 & 13.07 & 24.18 & 58.84 \\
Incidental Discovery           & 21 & \textbf{45.0} & 19.3 & 12.4 & 12.0 & 6.6 & 4.1 & 0.7 & 25.31 & 20.61 & 59.21 \\
Digital Channel Detection      & <1 & 31.5 & 2.2 & \textbf{51.4} & 6.7 & 1.5 & 6.6 & 0.2 & \textbf{38.90} & 15.29 & 53.46 \\
Survey-Based Discovery         & <1 & 9.2 & 1.0 & 0.6 & \textbf{83.1} & 4.9 & 0.8 & 0.4 & 29.41 & \textbf{13.72} & \textbf{65.00} \\
Institutional \& Community Reporting & 0 & -- & -- & -- & -- & -- & -- & -- & -- & -- & -- \\
Device Detection               & 0  & -- & -- & -- & -- & -- & -- & -- & -- & -- & -- \\
\hline
\multicolumn{12}{c}{\textbf{TFA Support}} \\
\hline
Digital Support Infrastructure & 62 & \textbf{45.6} & 19.2 & 11.5 & 12.5 & 6.4 & 4.2 & 0.6 & 24.04 & 21.07 & 59.70 \\
Technology Security Support    & 8  & \textbf{40.4} & 13.8 & 14.4 & 16.3 & 10.8 & 3.3 & 0.9 & 34.38 & 16.38 & 58.13 \\
Institutional Support          & <1 & \textbf{39.7} & 24.7 & 15.9 & 11.7 & 4.7 & 2.8 & 0.7 & 33.19 & 17.23 & 58.21 \\
Psychological Support          & <1 & \textbf{43.2} & 6.6 & 32.1 & 9.3 & 6.4 & 2.0 & 0.5 & 28.69 & 16.38 & 50.73 \\
Legal Support                  & 0 & -- & -- & -- & -- & -- & -- & -- & -- & -- & -- \\
Informal Support Networks      & <1 & \textbf{81.2} & 11.1 & 1.4 & 2.0 & 1.6 & 2.5 & 0.2 & \textbf{55.48} & \textbf{14.32} & \textbf{66.50} \\
\hline
\end{tabular}
\end{adjustbox}
\caption{Summary of TFA web content across taxonomy dimensions, showing category coverage in the corpus, distributions of dominant predicted emotions, and average readability scores. Higher Flesch and accessibility values indicate easier text, while higher ARI reflects more complex text; ``--'' denotes categories with no classified documents.}
\vspace{-8mm}
\label{tab:tfa-analysis}
\end{table*}

%% file: sections/04-conclusion.tex
\section{Conclusion}
\label{sec:conclusion}

This paper proposed a unified taxonomy of TFA, synthesising peer-reviewed studies into four dimensions (types, prevention, detection, support). The taxonomy was operationalised through a large-scale audit of 306 Australian institutional domains and 52,605 web pages using hierarchical zero-shot classification, emotion analysis, and readability profiling. Our results show that institutional content concentrates on harassment, comments abuse, and sexual abuse, while covert, privacy related, and economic abuse together account for fewer than 1\% of pages; that prevention and detection information is heavily skewed toward legal framing and incidental discovery, with proactive, technology based safeguards and reporting infrastructures rarely described; and that support is dominated by digital information hubs, with comparatively little web visible guidance into counselling, legal aid, or community based services, often expressed at late secondary or tertiary reading levels. Future work should apply this audit framework to other jurisdictions, examine how these informational gaps shape survivors' help-seeking and decision-making, and build taxonomy-aligned tools that surface clearer, more accessible, and actionable support pathways.

%% file: tables/co-occurence.tex
\begin{table}[h!]
\centering
\begin{adjustbox}{center,width=0.82\linewidth}
\begin{tabular}{l|l|c}
\hline
\textbf{Category} & \textbf{Subcategory} & \textbf{Count} \\
\hline

\multicolumn{3}{c}{\textbf{TFA Types}}\\\hline
Comments Abuse & Public Shaming & 11093 \\
Harassment & Threats of Eviction & 10297 \\
Sexual Abuse & Recorded Sexual Assault & 4310 \\
\hline

\multicolumn{3}{c}{\textbf{TFA Prevention}}\\\hline
Legal Support & Criminal Sanctions & 21344 \\
Family Safeguarding & Family-Based Early Detection & 7200 \\
Community-Based Prevention & Resource Provision and Navigation & 3433 \\
\hline

\multicolumn{3}{c}{\textbf{TFA Detection}}\\\hline
Clinical Detection & In-Person Professional Assessment & 21019 \\
Incidental Discovery & Unexpected Device Discovery & 10411 \\
Digital Channel Detection & Virtual Security Consultations & 84 \\
\hline

\multicolumn{3}{c}{\textbf{TFA Support}}\\\hline
Digital Support Infrastructure & Online Support Communities & 25912 \\
Technology Security Support & Security Clinics & 3917 \\
Institutional Support & Survivors Assistance Program & 334 \\
\hline
\end{tabular}
\end{adjustbox}
\caption{Top co-occurring category pairs for each TFA taxonomy.}
\label{tab:tfa-cooccurrence}
\end{table}


%% file: tables/support-examples.tex
\begin{table}[h!]
\centering
\begin{adjustbox}{center,width=0.88\linewidth}
\begin{tabular}{%
    p{3cm}<{\raggedright\arraybackslash}|
    p{6cm}<{\raggedright\arraybackslash}}
\hline
\textbf{Support Category} & \textbf{Example Domains} \\
\hline
Digital Support Infrastructure 
    & \url{southsafe.org.au}; \url{fcfcoa.gov.au}; \url{ndiscommission.gov.au} \\
\hline
Technology Security Support
    & \url{southsafe.org.au}; \url{gcyp.sa.gov.au}; \url{legalaid.vic.gov.au} \\
\hline
Institutional Support
    & \url{lwb.org.au}; \url{legalaid.vic.gov.au}; \url{nationallegalaid.org} \\
\hline
Psychological Support
    & \url{legalaid.wa.gov.au}; \url{headtohealth.gov.au}; \url{medicarementalhealth.gov.au} \\
\hline
Legal Support
    & \url{legalaid.wa.gov.au}; \url{legalaid.vic.gov.au}; \url{nationallegalaid.org} \\
\hline
Informal Network Support
    & \url{southsafe.org.au}; \url{headtohealth.gov.au} \\
\hline
\end{tabular}
\end{adjustbox}
\caption{Example Australian domains that provide different types of Technology-Facilitated Abuse (TFA) support, showing which organisations offer which classes of assistance.}
\label{tab:support-examples}
\end{table}